\documentclass{article}
\usepackage{spconf,amsmath,graphicx,hyperref}
\usepackage{amssymb}
\usepackage{booktabs}
\usepackage{graphicx}
\usepackage{lscape}
\usepackage{multirow}

\title{Harmonica: Accurate and Lightweight \\ Instrument-Agnostic Music Transcription}

\name{Longshen Ou, H\'ector Martel, Joe Hennessy-Priest, Taemin Cho}
\address{BandLab Technologies\\ 56 Neil Road, Singapore}

\begin{document} 
\ninept
 
\maketitle

\begin{abstract}
\label{abstract}
This paper introduces \textit{Harmonica}, a family of instrument-agnostic music transcription models built around multi-depth harmonic convolution. At each model scale, \textit{Harmonica} achieves the best performance among the evaluated models: the x-large model attains state-of-the-art performance in instrument-agnostic transcription, while the medium variant offers competitive accuracy with faster inference than all baselines. Pushing the limit of computational efficiency, the nano variant has only 26.3K parameters and runs at 1,622.5$\times$ real time, yet achieves a frame F1 of 0.796 on the development set, 14.6 percentage points higher than Basic Pitch. We further demonstrate that multi-depth harmonic convolution effectively exploits harmonic information to benefit transcription performance through comparative experiments with existing harmonic aggregation methods, including harmonic stacking, harmonic attention, single-depth harmonic convolution, and the HD-Conv layer.\footnote{Transcription examples and additional details on experiments and on-device deployment are available at \href{https://www.oulongshen.xyz/amt}{\nolinkurl{www.oulongshen.xyz/amt}}.}
 
\end{abstract}
\begin{keywords}
Automatic Music Transcription, Harmonic Convolution, Low-Resource, Instrument-Agnostic
\end{keywords}
\section{Introduction}
\label{sec:intro}

Automatic Music Transcription (AMT) is a fundamental problem in the music information retrieval community \cite{benetos2018automatic}, with decades of research devoted to improving transcription accuracy. However, most existing approaches are developed and evaluated under instrument-specific settings \cite{hawthorne2017onsets, kongHighResolutionPianoTranscription2021, semi-crf, ouExploringTransformersPotential2022a, hppnet, hft, triad, transkun, sft, rileyHighResolutionGuitar2024}, limiting their applicability to real-world scenarios involving diverse musical instruments. This raises a fundamental question: can a single model accurately transcribe performances across a wide range of pitched instruments?

This question remains largely underexplored, as attention has shifted instead to multi-instrument transcription, i.e., transcribing recordings in which several instruments play together \cite{mt3, perceivertf, ymt3, muscriptor}. However, whether such models reliably transcribe individual instruments across diverse types remains unclear. Arguably, strong instrument-agnostic performance is a prerequisite for robust multi-instrument transcription, which must additionally attribute each note to its source. A model that cannot transcribe instruments in isolation has little basis for handling mixtures of them reliably.


\begin{figure}[t]
\includegraphics[width=\columnwidth]{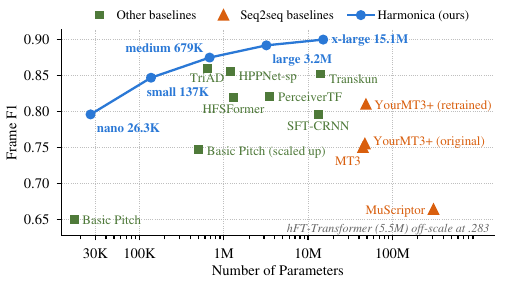}
\caption{Frame F1 versus model size on the held-out multi-source development set.}
\label{fig:res}
\end{figure}

Computational efficiency poses a further challenge. Recent AMT systems adopted increasingly large models for better accuracy \cite{muscriptor}, limiting their deployment on devices with constrained computation and memory. It also remains unclear which architectural paradigm benefits most from scaling: attention-based and sequence-to-sequence models are computationally expensive, yet do not consistently deliver strong transcription performance, as our experiments show. Can a better architecture push the accuracy-efficiency frontier forward, and scale more rewardingly than sequence-to-sequence models?

In this work, we first benchmark recent AMT models under a common multi-source, instrument-agnostic evaluation protocol to assess their ability to generalize across diverse pitched instruments. We then propose \textit{Harmonica}, a family of instrument-agnostic music transcription models built around multi-depth harmonic convolutions. The family spans a wide range of computational budgets, with even its smallest variant maintaining competitive accuracy. Our contributions are summarized as follows:

\label{contribution}
\begin{itemize}
\item We show that multi-depth harmonic convolution exploits harmonic information  for  AMT performance more effectively than existing harmonic aggregators such as harmonic stacking, single-depth harmonic convolution, and harmonic attention. Further, we introduce a parameter-efficient formulation of harmonic convolution that matches the performance of HD-Conv \cite{hppnet} while reducing harmonic-layer parameters by 28.5\%.
\item At each model scale, \textit{Harmonica} achieves the best overall performance among the evaluated models. The x-large model attains state-of-the-art accuracy; the medium variant achieves near-SOTA accuracy while running faster than every baseline; and the nano variant, at a similar parameter scale to Basic Pitch, is substantially more accurate while running more than 4$\times$ faster.
\item We retrain various AMT models under a common multi-source, instrument-agnostic setting, yielding a comprehensive benchmark of instrument-agnostic music transcription performance on eight test sets spanning diverse instrument properties.
\end{itemize}

\section{Related Works}
\label{sec:related}


\smallskip\noindent\textbf{Instrument-Agnostic Music Transcription.}
Basic Pitch \cite{basic-pitch}, a prominent instrument-agnostic AMT system, uses a lightweight convolutional neural network (CNN) to transcribe a wide range of pitched instruments, though its accuracy leaves substantial room for improvement. HFSFormer \cite{hfsformer} addresses the same setting with a CNN-Transformer architecture, reporting strong results on MAESTRO and GuitarSet. Beyond these, most recent work targets either a single instrument or full multi-instrument mixtures, so how existing architectures generalize across instrument families is largely unstudied.

\smallskip\noindent\textbf{Sequence-to-Sequence Models.}
Hawthorne et al. \cite{DBLP:conf/ismir/HawthorneSSME21} introduced the sequence-to-sequence formulation to piano AMT with a T5-style architecture. MT3 \cite{mt3} extended this paradigm to multi-instrument transcription, followed by YourMT3 \cite{ymt3} and MIROS \cite{chaturvediAdvancingMultiInstrumentMusic2026}, and MuScriptor \cite{muscriptor} scaled it to over one billion parameters. Without explicit musical priors, these models learn the mapping from acoustic features to symbolic notation entirely from data, and therefore demands large models and large training sets. The autoregressive decoder benefits onset prediction by inferring note location from context, yet offset detection depend on grounded analysis of sustained acoustic evidence rather than on sequence structure and remain difficult. 

\smallskip\noindent\textbf{Frame-Level Models.}
Frame-level prediction is a more efficient alternative and has been studied extensively in AMT. Onsets and Frames \cite{hawthorne2017onsets} predicts note onsets and frames for piano transcription with a convolutional and recurrent network. More recently, hFT-Transformer \cite{hft} employs a frequency-time Transformer encoder for frame-level prediction of note attributes, while PerceiverTF \cite{perceivertf} uses a Perceiver architecture for multi-instrument transcription. These models do not need a heavy autoregressive decoder, but still rely on substantial capacity to learn the mapping from acoustic features to musical structure: in our experiments, hFT-Transformer underperforms our nano model despite having 200$\times$ more parameters. Other work introduces musical structure through event priors such as semi-CRFs \cite{semi-crf, transkun}, or sharpens temporal localization through boundary regression \cite{kongHighResolutionPianoTranscription2021} or an optimal transport loss \cite{sft}, but their effectiveness for instrument-agnostic AMT remains unclear.

\smallskip\noindent\textbf{Harmonic Aggregation.}
The efficiency of frame-level models can be further improved by aggregating information across harmonically related frequency bins inside the network. HPPNet \cite{hppnet} uses Harmonic Dilated Convolution (HD-Conv) and outperforms Onsets and Frames with nearly 1/20 of its parameters. TriAD \cite{triad} improves parameter efficiency further with a 3-D harmonic layer over octave-folded time-frequency features, and SFT-CRNN \cite{sft} employs an attention-based harmonic aggregator to strengthen a CRNN backbone. HFSFormer \cite{hfsformer} applies frequency-separable transformer blocks to harmonically aggregated features. Harmonic aggregation also enables extremely lightweight models by extending the convolutional receptive field along the frequency axis in shallow networks, including HPPNet-tiny at 151K parameters with HD-Conv and Basic Pitch at 16.8K parameters with harmonic stacking.

However, it remains unclear which harmonic aggregation strategy is most effective for instrument-agnostic transcription. Further, convolutional approaches tend to aggregate harmonic information only once, either at the input through harmonic stacking or in a single harmonic convolution layer, despite the complex interactions among harmonics and subharmonics that arise when several notes sound at once with overlapping harmonic series. This motivates exploring harmonic convolution at multiple levels of abstraction.


\section{Method}
\label{sec:model}

\subsection{Problem Formulation}

Let $x \in \mathbb{R}^{L}$ be an audio signal of $L$ samples, sampled at a rate $f_s$, containing a single instrument with unknown identity. Instrument-agnostic music transcription aims to recover a set of note events $y = \{(s_i, e_i, p_i)\}_{i=1}^{N}$, where $s_i < e_i$ are the onset and offset times in seconds, $p_i$ is a MIDI pitch from $\mathcal{P} = \{21, \dots, 108\}$, and the note count $N$ is unknown \textit{a priori}. Notes may overlap in time, except when they share a pitch.

A fixed time-frequency transform $\Phi$ maps $x$ to $X = \Phi(x) \in \mathbb{R}^{F \times T}$, with $F$ frequency bins and $T = \lceil L/h \rceil$ frames at hop size $h$. A learned network $f_\theta$ predicts three frame-level matrices, 
\begin{equation}
  \bigl(\hat{Y}^{\mathrm{on}}, \hat{Y}^{\mathrm{sus}}, \hat{Y}^{\mathrm{off}}\bigr) = f_\theta(X), \qquad \hat{Y}^{(\cdot)} \in [0,1]^{|\mathcal{P}| \times T},
\end{equation}
denoting onset, sustain, and offset activation. The temporal resolution is inherited from $\Phi$, whereas the frequency axis is reduced to one bin per semitone, aligning with the target's pitch axis $\mathcal{P}$. A deterministic decoding algorithm $g$ recovers the note events, $\hat{y} = g(\hat{Y}^{\mathrm{on}}, \hat{Y}^{\mathrm{sus}})$, by thresholding onset peaks and tracking the sustain until it falls below a separate threshold. Note that $\hat{Y}^{\mathrm{off}}$ serves as auxiliary supervision only and does not participate in decoding. 

\subsection{Spectrogram Representation}

Many AMT systems operate on the STFT spectrogram \cite{perceivertf, ymt3} or one of its variants \cite{kongHighResolutionPianoTranscription2021, mt3}, which offers limited frequency resolution in the low register: a 2048-sample window on 16~kHz audio yields a bin spacing of 7.8~Hz, comparable to the interval between C3 (130.8~Hz) and C$\sharp$3 (138.6~Hz), so the fundamentals of adjacent lower notes are difficult to resolve. We therefore adopt the CQT spectrogram \cite{brown1991calculation}, whose logarithmically spaced bins provide constant resolution in semitones across the full pitch range. The input axis is thus aligned with the output pitch axis, so the model need not learn a mapping from frequency to pitch, nor infer fundamentals in regions where they are unresolvable. Instead, it can focus on an in-place mapping, namely attributing spectral energy to the correct fundamental frequency bin. This simplifies the learning problem and enables building parameter-efficient models \cite{hppnet, basic-pitch, hfsformer}.


\subsection{Model}

\begin{figure}[tb]
\includegraphics[width=\columnwidth]{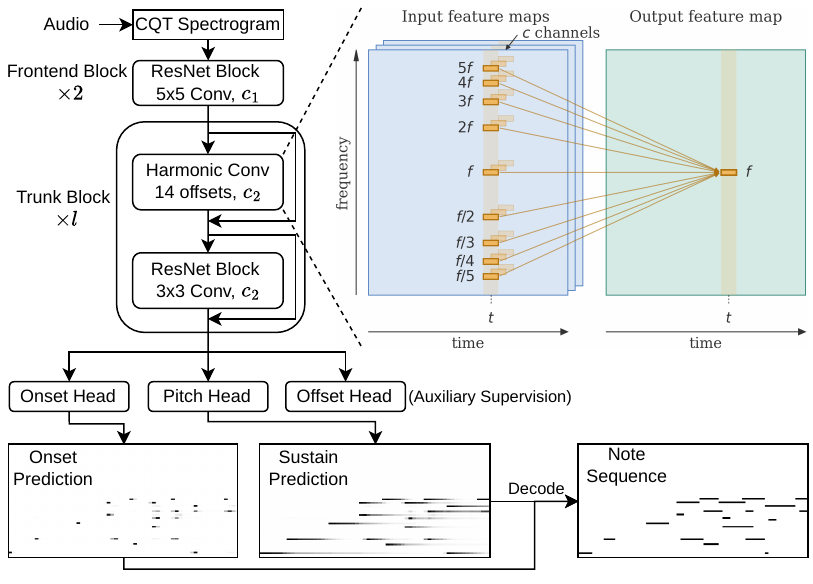}
\caption{Model architecture of \textit{Harmonica}.  }
\label{fig:arch}
\end{figure}

The architecture is shown in Figure~\ref{fig:arch}. The input is a CQT spectrogram calculated on 16~kHz audio, with three bins per semitone and a 32~ms frame interval, first processed by a shallow frontend of two ResNet blocks \cite{he2016deep}, each with two $5 \times 5$ convolutions and channel size $c_1$. Then, a strided convolution with a $3 \times 1$ kernel and stride $3$ along the frequency axis expands the channel size to $c_2$ and downsamples the frequency axis to one bin per semitone. 

The frontend is followed by $l$ trunk blocks, each pairing a harmonic convolution layer with a ResNet block. The harmonic convolution layer aggregates information accross harmonically related frequency bins within each frame, followed by normalization, a residual connection, and ReLU activation. The ResNet block contains two $3 \times 3$ convolutions and a residual connection. Stacking these blocks yields a deep trunk in which harmonic aggregation is performed at multiple levels of abstraction. Then, the trunk output is fed to three heads, one for each of the onset, sustain, and offset targets. Each head consists of a single-layer frequency-grouped LSTM (FGLSTM) \cite{hppnet} followed by a linear projection producing one scalar per time-pitch position.

\subsection{Harmonic Convolution}

Harmonic convolution refines the hypothesis that a given frequency bin corresponds to a fundamental, based on the energy at the harmonic and subharmonic positions of that bin within the same frame. Energy at the harmonic positions supports the hypothesis, whereas energy at the subharmonic positions suggests the bin may itself be a harmonic of a lower fundamental. Because these positions are spaced unevenly along the log-frequency axis, the operation cannot be realized by a single dilated convolution.

The Harmonic Dilated Convolution (HD-Conv) \cite{hppnet} realizes harmonic convolution with a set of $3 \times 1$ convolutional branches, each dilated with a different interval to capture one harmonic and its corresponding subharmonic. The branch outputs are then summed. This design is not parameter-efficient: the center bin is covered by all branches and thus redundantly parameterized once per branch.

We remove this redundancy with a shift-and-aggregate formulation. The feature map is shifted along the frequency axis by a set of offsets $\{o_1, \dots, o_M\}$, each indicating the distance in semitones between a pitch and one of its harmonics or subharmonics, and the shifted copies are concatenated channel-wise, then aggregated by a single $1 \times 1$ convolution mapping $c_2 M$ channels to $c_2$. Each harmonic position is thus parameterized exactly once. For the configuration used in our models (7 harmonics and 7 subharmonics), this reduces the parameter count of the harmonic layers by 28.5\% relative to HD-Conv while preserving performance, as shown in Table~\ref{tab:comparative}.
 
\section{Experiments}
\label{sec:experiments}


\begin{table*}[tb]
\centering
\caption{Performance comparison of various models   on instrument-agnostic music transcription. \textbf{\#Param}: number of parameters; \textbf{xRT}: inverse real-time factor of inference speed; \textbf{OnP}: note F1 considering onset and pitch; \textbf{Frm}: frame-level F1 score. }
\label{tab:main}
\resizebox{\textwidth}{!}{%
\begin{tabular}{@{}lrrcccccccccccccccc@{}}
\toprule
                                   & \multicolumn{1}{l}{} & \multicolumn{1}{l}{}             & \multicolumn{2}{c}{\textbf{MAESTRO}} & \multicolumn{2}{c}{\textbf{MAPS}} & \multicolumn{2}{c}{\textbf{GuitarSet}} & \multicolumn{2}{c}{\textbf{GAPS}} & \multicolumn{2}{c}{\textbf{EGDB}} & \multicolumn{2}{c}{\textbf{GOAT}} & \multicolumn{2}{c}{\textbf{URMP-stem}} & \multicolumn{2}{c}{\textbf{Slakh-stem}} \\ \cmidrule(l){4-19} 
\multicolumn{1}{c}{\textbf{Model}} & \textbf{\#Param}     & \multicolumn{1}{c}{\textbf{xRT}} & \textbf{OnP}      & \textbf{Frm}     & \textbf{OnP}    & \textbf{Frm}    & \textbf{OnP}       & \textbf{Frm}      & \textbf{OnP}    & \textbf{Frm}    & \textbf{OnP}    & \textbf{Frm}    & \textbf{OnP}    & \textbf{Frm}    & \textbf{OnP}       & \textbf{Frm}      & \textbf{OnP}       & \textbf{Frm}       \\ \midrule

MT3 \cite{mt3}                               & 44.7M                & 68.7x                            & .958              & .890             & .801            & .732            & .891               & .883              & .659            & .616            & .448            & .448            & .291            & .398            & .829               & .879              & .900               & .826               \\
YourMT3+ \cite{ymt3}                           & 48.5M                & 45.6x                            & .957              & .769             & .828            & .691            & .889               & .835              & .917            & .748            & .827            & .725            & .680            & .703            & .926               & .920              & .896               & .753               \\
MuScriptor \cite{muscriptor}                        & 306M               & 17.0x                            & .766              & .625             & .806            & .683            & .808               & .732              & .781            & .742            & .789            & .731            & .710            & .783            & .711               & .835              & .457               & .488               \\
hFT-Transformer \cite{hft}                   & 5.5M                 & 91.7x                            & .573              & .283             & .642            & .453            & .791               & .803              & .775            & .648            & .707            & .653            & .664            & .681            & .802               & .854              & .624               & .637               \\
PerceiverTF \cite{perceivertf}                       & 3.5M                & 152.7x                           & .819              & .816             & .728            & .678            & .853               & .865              & .843            & .708            & .799            & .737            & .618            & .697            & .808               & .860              & .751               & .760               \\
Transkun \cite{transkun}                          & 14.1M                & 59.2x                            & .926              & .874             & .833            & .772            & .873               & .878              & .905            & .764            & .820            & .726            & .638            & .745            & .921               & .933              & .833               & .810               \\
SFT-CRNN \cite{sft}                          & 13.2M                & 301.5x                           & .964              & .836             & .877            & .583            & .898               & .767              & .938            & .415            & .874            & .630            & .746            & .667            & \textbf{.933}      & .898              & .893               & .754               \\
Basic Pitch \cite{basic-pitch}                        & 16.8K                & 392.7x                           & .578              & .577             & .597            & .621            & .776               & .827              & .712            & .671            & .718            & .658            & .553            & .647            & .687               & .894              & .394               & .502               \\
HPPNet-sp \cite{hppnet}                         & 1.2M                 & 154.5x                           & .956              & .892             & \textbf{.882}   & \textbf{.825}   & .893               & .875              & .929            & .793            & .857            & .749            & .739            & .770            & .911               & .909              & .870               & .814               \\
TriAD \cite{triad}                             & 635K                 & 256.4x                           & .934              & .879             & .863            & .794            & .889               & .874              & .921            & .782            & .857            & .756            & .706            & .772            & .894               & .892              & .836               & .802               \\
HFSFormer \cite{hfsformer}                                      & 1.3M                 & 154.1x                           & .871              & .852             & .821            & .754            & .857               & .856              & .852            & .740            & .806            & .746            & .664            & .695            & .774               & .815              & .732               & .749               \\ \midrule
Ours, nano                         & 26.3K                & \textbf{1622.5x}                          & .880              & .831             & .817            & .760            & .863               & .852              & .887            & .733            & .805            & .725            & .686            & .738            & .796               & .833              & .731               & .712               \\
Ours, small                        & 137K                 & 1335.5x                          & .928              & .881             & .848            & .778            & .881               & .869              & .915            & .777            & .848            & .756            & .709            & .768            & .881               & .895              & .822               & .797               \\
Ours, medium                       & 679K                 & 848.5x                           & .951              & .902             & .866            & .790            & .896               & .881              & .930            & .799            & .869            & .764            & .720            & .796            & .924               & .930              & .877               & .842               \\
Ours, large                        & 3.2M                 & 346.5x                           & .962              & .915             & .870            & .790            & .907               & .895              & .937            & .814            & .876            & .766            & .747            & .807            & .926               & .930              & .905               & .865               \\
Ours, x-large                      & 15.1M                & 123.0x                           & \textbf{.968}     & \textbf{.922}    & .874            & .785            & \textbf{.909}      & \textbf{.898}     & \textbf{.939}   & \textbf{.820}   & \textbf{.881}   & \textbf{.773}   & \textbf{.750}   & \textbf{.809}   & .925               & \textbf{.934}     & \textbf{.918}      & \textbf{.877}      \\ \bottomrule
\end{tabular}
}
\end{table*}

\subsection{Experiment Settings}

We adopt eight datasets: MAESTRO \cite{maestro}, MAPS \cite{maps}, GuitarSet \cite{guitarset}, GAPS \cite{gaps}, EGDB \cite{egdb}, GOAT \cite{goat}, URMP \cite{urmp}, and Slakh2100 \cite{slakh}. For URMP and Slakh2100, we adopt the track-wise audio--MIDI pairs, excluding the mixture audio. The Disklavier subset of MAPS serves as an out-of-domain test set, while the remaining datasets are used for both training and testing. Following MT3 \cite{mt3}, multi-source batches are formed by temperature sampling: training examples are 5-second audio clips, and each clip is drawn from source $i$ with probability proportional to $(n_i / \sum_j n_j)^{\alpha}$, where $n_i$ is the number of hours in source $i$. We set $\alpha = 0.5$, which upsamples the smaller sources so that they are not overwhelmed by the larger ones. The official validation splits are merged into the training set, as they are too large to evaluate frequently and their source distribution does not reflect the temperature-sampled training distribution. We instead hold out 4.3 hours of audio under the same temperature sampling as a development set for monitoring training progress and determining optimal model-specific thresholds.

We compared against eleven recent music transcription models, covering sequence-to-sequence models (MT3 \cite{mt3}, YourMT3+ \cite{ymt3}, MuScriptor \cite{muscriptor}), general frame-level models (hFT-Transformer \cite{hft}, PerceiverTF \cite{perceivertf}, Transkun \cite{transkun}, SFT-CRNN \cite{sft}), and models with harmonic aggregation (Basic Pitch \cite{basic-pitch}, HPPNet \cite{hppnet}, TriAD \cite{triad}, HFSFormer \cite{hfsformer}). Eight of them are retrained under our multi-source instrument-agnostic protocol. The remaining three are evaluated using their released checkpoints. Basic Pitch and MT3 are excluded because their training data substantially overlaps with ours, and MuScriptor because its pipeline combines synthetic-data pretraining with reinforcement learning post-training, which is not aligned with our supervised setup; we adopt its medium checkpoint.

We report note-level and frame-level F1 scores. Two note-level F1 scores from \cite{raffel2014mir_eval} are used: \textbf{OnP} counts a predicted note as correct when its onset and pitch both match those of a reference note, whereas \textbf{OnPOff} additionally requires a matching offset. The onset tolerance is 50~ms, and the offset tolerance is $\max(0.2\,d, 50~\mathrm{ms})$ for a reference note of duration $d$. Since the baselines operate at different temporal resolutions, we derive frame-level predictions from the predicted note events and compute frame-level F1, denoted as \textbf{Frm}, using a fixed frame size of 25~ms for fair comparison. We additionally report inference speed as the inverse real-time factor, denoted as \textbf{xRT}, measured on an RTX 4090 GPU.

For the proposed architecture, varying $l$, $c_1$, and $c_2$ yields models at different parameter scales. From nano to x-large, we set $(c_2, l)$ to $(16, 2)$, $(32, 3)$, $(64, 4)$, $(128, 5)$, and $(256, 6)$, with $c_1 = c_2/4$ in every configuration. Instance normalization is adopted rather than batch normalization throughout the network. Our models are trained with summed binary cross-entropy loss from onset, sustain, and offset, for 100k steps with a batch size of 8, using the AdamW optimizer \cite{loshchilov2017decoupled}, a cosine learning rate schedule, and an initial learning rate of $2 \times 10^{-4}$. YourMT3+ and PerceiverTF are trained for twice as many steps to reach convergence, and hFT-Transformer for ten times as many because of its slow convergence. All remaining baselines are trained under the same computational budget as ours.

\subsection{Main Result}

The benchmark results are reported in Table~\ref{tab:main}. Our models deliver the best overall performance at every scale. (1) The x-large model ranks first in frame F1 on all in-domain test sets, and first in OnP F1 as well except on URMP-stem, where SFT-CRNN is slightly ahead. (2) On the in-domain test sets, the medium model performs comparably to the best baseline in every column while running faster than all baselines, at 848.5$\times$ real time. (3) The nano model, with only 26.3K parameters at 1,622.5$\times$ real time, performs comparably to PerceiverTF using 1/133 of the parameters and running over ten times faster, and outperforms Basic Pitch by a wide margin on all test sets other than URMP. (4) Accuracy improves consistently with scale on the held-out development set, where every configuration from medium upward surpasses all baselines in frame F1, defining a new accuracy-efficiency frontier (Figure~\ref{fig:res}). Taken together, these results demonstrate the effectiveness of the proposed architecture for instrument-agnostic music transcription.

The baselines also reveal several patterns. YourMT3+ retrained under our protocol reaches competitive OnP F1 but falls behind on frame F1, which is consistent with the difficulty of predicting offsets for sequence-to-sequence models discussed in Section~\ref{sec:related}. Performance of MT3 also degrades sharply on GOAT, the distorted electric guitar set, attaining an OnP F1 of only .291. Similar note--frame asymmetry exists for SFT-CRNN for a different reason: many detected onsets receive no matching offset and the corresponding notes are left ringing, producing abnormally long durations that depress frame F1 to .415 on GAPS despite an OnP F1 of .938. On the out-of-domain piano test set MAPS, HPPNet-sp performs best, which is expected given that its architecture is designed specifically for piano. 

\begin{figure}[tb]
\centering
\includegraphics[width=\columnwidth]{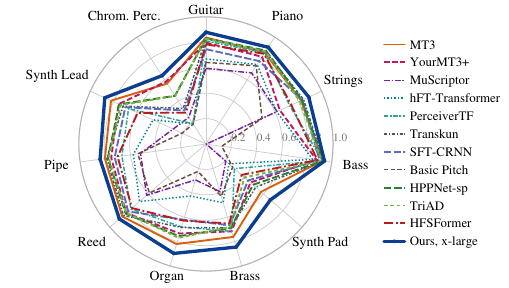}
\caption{Frame F1 per instrument family on Slakh2100.}
\label{fig:perf_inst}
\end{figure}

Figure~\ref{fig:perf_inst} breaks performance down by instrument family. Our x-large model is the strongest on every family, and its scores are more evenly distributed across families. On the other hand, many baselines such as PerceiverTF and HFSFormer degrade sharply on the harder families, particularly chromatic percussion and synth pad. This suggests that the proposed architecture generalizes well across diverse pitched instruments.

\subsection{Comparative Studies}

\begin{table}[tb]
\centering
\caption{Comparative study of different harmonic aggregators. \textbf{OnPOff}: note F1 score considering onset, pitch, and offset. All experiments are conducted three times under different random seeds, and the average performance is reported.}
\label{tab:comparative}
\resizebox{\columnwidth}{!}{%
\begin{tabular}{llccc}
\hline
\multicolumn{1}{c}{\textbf{Harmonic Aggregation}} & \textbf{Depth}        & \textbf{OnP} & \textbf{OnPOff} & \textbf{Frm} \\ \hline
No harmonic aggregation                          & \multicolumn{1}{c}{-} & .822         & .635            & .823         \\
Harmonic stacking \cite{basic-pitch}              & Single                & .851         & .688            & .848         \\
Harmonic attention \cite{sft}                     & Multi.                & .877         & .737            & .868         \\
Conv, w/ shift-and-agg.     & Single                & .877         & .731            & .867         \\
Conv, w/ shift-and-agg. (Ours)                    & Multi.                & .889         & .745            & .876         \\ \hline
Conv, w/ HD-Conv \cite{hppnet}                    & Multi.                & .886         & .745            & .877         \\ \hline
\end{tabular}%
}
\end{table}

Table~\ref{tab:comparative} compares harmonic aggregation strategies, all evaluated with variants of our medium model so that only the aggregator differs, with metrics computed on the held-out development set. Removing every harmonic layer costs 6.7 points of OnP and 11.0 points of OnPOff relative to our model, confirming that harmonic aggregation is essential for transcription performance. Among the aggregators, harmonic stacking at the input, as used in Basic Pitch, is the weakest, and harmonic attention performs on par with single-depth harmonic convolution on OnP and Frm. All three remain below our multi-depth harmonic convolution, which improves OnPOff by 1.4 points over the single-depth variant. This supports our central claim that harmonic convolution is more effective when applied at multiple levels of abstraction than when confined to one. Finally, replacing our shift-and-aggregate formulation with HD-Conv in every harmonic layer yields essentially the same scores, at the cost of 39.9\% more parameters per harmonic layer.

\section{Conclusion}
\label{sec:conclusion}

We benchmarked recent AMT models under a common multi-source, instrument-agnostic protocol and showed that existing approaches can vary substantially in their ability to generalize across instruments. We introduced \textit{Harmonica}, a family of efficient convolution-based models that scales well across computational budgets and consistently outperforms prior models at comparable scales. Our analysis further showed that multi-depth harmonic convolution provides an effective and parameter-efficient way to exploit harmonic information. Together, these results demonstrate the potential of efficient, instrument-agnostic architectures for practical AMT, with \textit{Harmonica} already deployed as the Audio-to-MIDI service in BandLab to support accessible music creation.

\vfill\pagebreak


\footnotesize
\bibliographystyle{IEEEbib}
\bibliography{strings,refs}
\ninept
\appendix
\section{Efficiency of On-Device Deployment}

To illustrate the computational efficiency of our model, we report the inference speed and peak memory consumption of the medium model deployed on an iPhone 15 Pro Max, transcribing Tommy Emmanuel's \textit{Angelina} (156~s in duration). With ONNX Runtime on the CPU, inference takes 22.16~s at 7.0$\times$ real time, with a peak memory consumption of 382.6~MB. With LiteRT on the GPU, inference takes 4.86~s at 32.1$\times$ real time, with a peak memory consumption of 445.0~MB. Note that in this experiment the model implementation is further optimized for memory consumption and inference latency, while the architecture and parameter count remain the same as those described in the main paper.

For context, the Basic Pitch paper \cite{basic-pitch} reports the following figures for transcribing long audio on a 2017 MacBook Pro with an Intel Core i7 CPU: MI-AMT \cite{wu2020multi} runs at 4.8$\times$ real time with a peak memory consumption of 3.3~GB, Basic Pitch at 19.4$\times$ real time with 951~MB, Onsets and Frames \cite{hawthorne2017onsets} at 5.4~GB, and VOCANO \cite{hsu2021vocano} at 8.5~GB. Notably, our medium model runs faster than Basic Pitch on mobile hardware despite having roughly 40 times more parameters, and its peak memory consumption stays below half of Basic Pitch's.

\section{Dataset Information}

This section provides further details on each dataset used in our experiments, its split configuration, and the rationale for including it. Table~\ref{tab:data} gives an overview of all datasets.

\smallskip\noindent\textbf{MAESTRO.} MAESTRO (MIDI and Audio Edited for Synchronous TRacks and Organization) \cite{maestro} is a common benchmark for training and evaluating piano transcription systems. We adopt version 3, which contains about 198.6 hours of piano performance recordings. The recordings were captured on a Yamaha Disklavier piano, ensuring accurate alignment between the audio and MIDI files.

\smallskip\noindent\textbf{MAPS.} MAPS (MIDI Aligned Piano Sounds) \cite{maps} is a piano dataset containing both synthesized and recorded piano sounds, including isolated notes, chords, and full pieces. Its two recorded subsets, ENSTDkAm and ENSTDkCl, contain full pieces played on two real Disklavier pianos and are commonly used as out-of-domain test sets, which is also how we use them in our benchmark.

\smallskip\noindent\textbf{GuitarSet.} GuitarSet \cite{guitarset} contains acoustic guitar improvisations over three chord progressions, performed in various keys, tempos, and musical genres by six players with more than ten years of experience. Each performance is recorded from both a microphone and an amplifier, and we treat the two as separate audio-MIDI pairs. Recording uses a hexaphonic pickup with one channel per string, so annotations can be obtained conveniently through monophonic audio analysis. Following MT3 \cite{mt3}, we use progressions 1 and 2 for training and progression 3 for testing. The provided JAMS annotations are converted to MIDI with REMI-z \cite{ou2026unifying}.

\smallskip\noindent\textbf{GAPS.} GAPS (Guitar-Aligned Performance Scores) \cite{gaps} is a classical guitar dataset providing audio-score pairs with high-resolution aligned MIDI. Classical guitar differs in tonal properties from acoustic guitar, as reflected in the sharp performance degradation of MT3 on this dataset, and it is a sufficiently important instrument to include in a real-world transcription benchmark.

\begin{table}[tb]
\centering
\caption{Dataset information. The statistics for the training split is calculated before holding out the development set.}
\label{tab:data}
\begin{tabular}{@{}clcrr@{}}
\toprule
\textbf{Purpose}        & \multicolumn{1}{c}{\textbf{Dataset}} & \textbf{Original Split} & \multicolumn{1}{c}{\textbf{\#Record}} & \multicolumn{1}{c}{\textbf{Hour}} \\ \midrule
\multirow{10}{*}{Train} & MAESTRO                              & Train                   & 962                                   & 159.2                             \\
                        & MAESTRO                              & Valid                   & 137                                   & 19.4                              \\
                        & GuitarSet                            & Train                   & 480                                   & 3.9                               \\
                        & GAPS                                 & Train                   & 270                                   & 14.8                              \\
                        & EGDB                                 & Train                   & 1290                                  & 9.7                               \\
                        & GOAT                                 & Train                   & 870                                   & 31.3                              \\
                        & URMP-stem                            & Train                   & 117                                   & 3.9                               \\
                        & Slakh-stem                           & Train                   & 13546                                 & 948.2                             \\
                        & Slakh-stem                           & Valid                   & 2833                                  & 195.1                             \\
                        & Total                                &                         & 20505                                 & 1385.5                            \\ \midrule
\multirow{8}{*}{Dev}    & MAESTRO                              & Train and valid         & 7                                     & 1.0                               \\
                        & GuitarSet                            & Train                   & 16                                    & 0.1                               \\
                        & GAPS                                 & Train                   & 4                                     & 0.3                               \\
                        & EGDB                                 & Train                   & 25                                    & 0.2                               \\
                        & GOAT                                 & Train                   & 11                                    & 0.4                               \\
                        & URMP-stem                            & Train                   & 3                                     & 0.1                               \\
                        & Slakh-stem                           & Train and valid         & 32                                    & 2.2                               \\
                        & Total                                &                         & 98                                    & 4.3                               \\ \midrule
\multirow{9}{*}{Test}   & MAESTRO                              & Test                    & 177                                   & 20.0                              \\
                        & MAPS                                 & Test                    & 60                                    & 4.4                               \\
                        & GuitarSet                            & Test                    & 240                                   & 2.2                               \\
                        & GAPS                                 & Test                    & 30                                    & 1.8                               \\
                        & EGDB                                 & Test                    & 150                                   & 1.1                               \\
                        & GOAT                                 & Test                    & 42                                    & 1.4                               \\
                        & URMP-stem                            & Test                    & 32                                    & 0.7                               \\
                        & Slakh-stem                           & Test                    & 1539                                  & 113.4                             \\
                        & Total                                &                         & 2270                                  & 145.0                             \\ \bottomrule
\end{tabular}
\end{table}

\smallskip\noindent\textbf{EGDB.} The variety of sound is part of what makes the guitar so versatile: different guitars produce entirely different timbres, and in some respects behave as distinct instruments sharing a common name. EGDB \cite{egdb} is an electric guitar dataset containing performances of 240 tablatures, each provided as a direct input signal together with five amplifier renderings, mostly in clean tone. Its tonal characteristics differ from those of the two guitar types described above, which is again reflected in the weak performance of MT3 on this dataset. As electric guitar is among the most widely used instruments in modern music, a transcription system intended for real-world use must handle it reliably.

\smallskip\noindent\textbf{GOAT.} Electric guitar is commonly played with distortion, a signature tone in many genres. Waveform clipping substantially alters the spectral characteristics of the signal and introduces a large number of additional harmonics, which poses a significant challenge for transcription systems. To test whether models can cope with such material, we adopt the GOAT dataset \cite{goat}, in which two thirds of the recordings are distorted electric guitar. It is the most difficult dataset in our benchmark.

\smallskip\noindent\textbf{URMP.} URMP \cite{urmp} contains recordings of 44 classical ensemble pieces, with separate audio-MIDI pairs for each track. The instruments covered are violin, viola, cello, double bass, trumpet, trombone, horn, tuba, flute, oboe, clarinet, saxophone, and bassoon, spanning the string, brass, and woodwind families. Following MT3, we use pieces 1, 2, 12, 13, 24, 25, 31, 38, and 39 for validation and the remainder for training. To distinguish this setting from experiments using the mixture audio, we denote it URMP-stem in Table~\ref{tab:main}.

\smallskip\noindent\textbf{Slakh.} Slakh \cite{slakh} is a synthesized multitrack dataset in the pop genre, whose MIDI sources are a subset of the Lakh MIDI Dataset \cite{raffel2016learning}. Each track is rendered with at least four instruments: drums, piano, bass, and guitar. It spans a wide range of instrument families across the General MIDI specification. The sound effects family is not rendered and is therefore excluded, leaving all 12 pitched families. Following \cite{perceivertf, ymt3}, we merge the Strings and String Ensemble families, yielding 11 families: Piano, Strings, Bass, Synth Pad, Brass, Organ, Reed, Pipe, Synth Lead, and Chromatic Percussion, ordered by the amount of data available. We adopt the track-wise audio-MIDI pairs of all pitched instruments and use neither the drum tracks nor the mixture audio. To distinguish this setting from experiments using the mixture audio, we denote it Slakh-stem in Table~\ref{tab:main}.

\smallskip\noindent\textbf{Held-Out Development Set.} This set is drawn from the training splits of all sources under the same temperature sampling used during training, giving 98 songs and 4.33 hours in total. The selected samples are excluded from training to prevent data leakage. It provides a single-number estimate of instrument-agnostic transcription performance, which supports checkpoint selection and threshold sweeping.

\begin{figure}[tb!]
\includegraphics[width=\columnwidth]{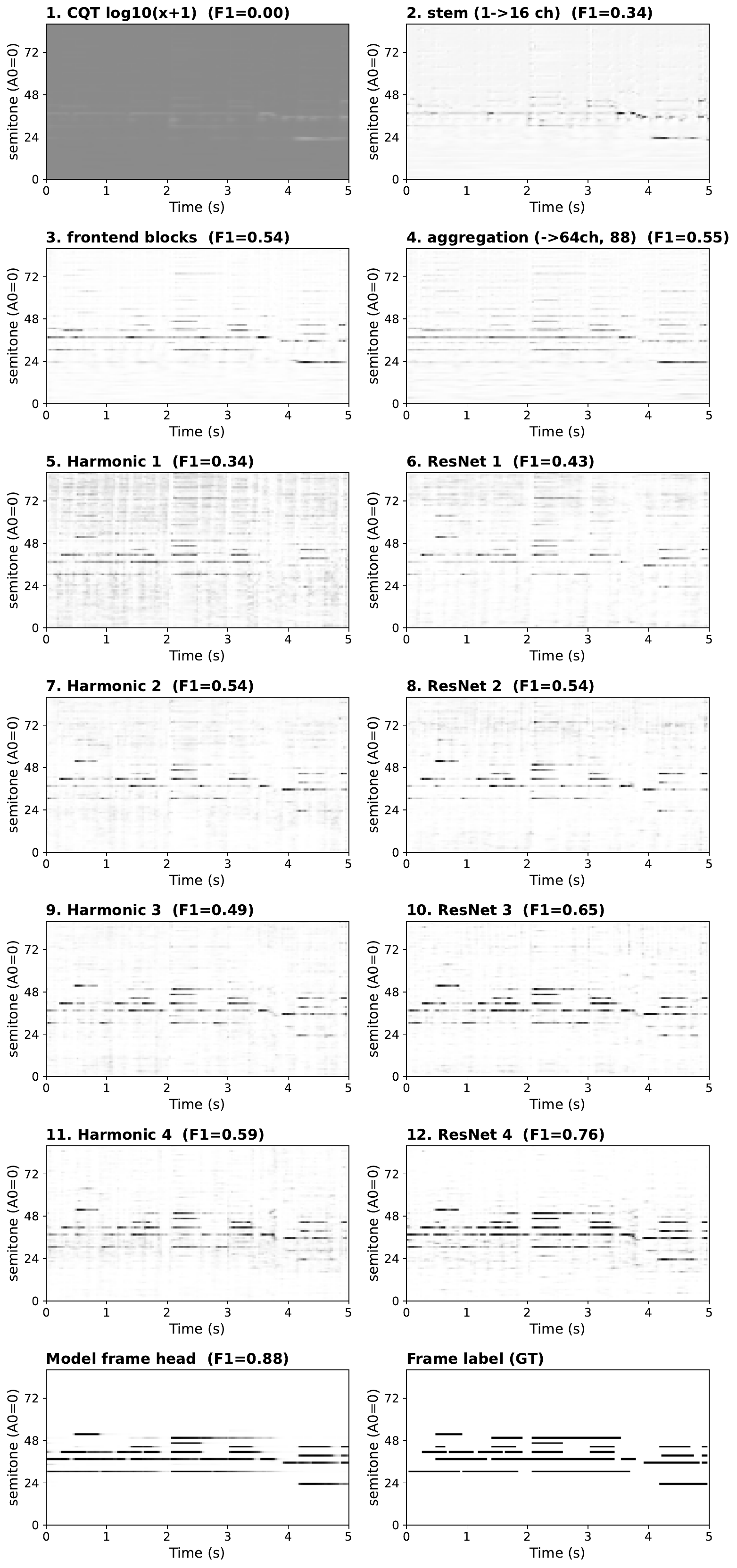}
\caption{Readouts from the linear probe of each layer's output.}
\label{fig:probe}
\end{figure}

\begin{table*}[tb!]
\centering
\caption{Full benchmark results 1/2.}
\label{tab:full1}
\resizebox{\textwidth}{!}{%
\begin{tabular}{@{}lrrcccccccccccc@{}}
\toprule
                                   & \multicolumn{1}{l}{} & \multicolumn{1}{l}{}             & \multicolumn{3}{c}{\textbf{MAESTRO}}            & \multicolumn{3}{c}{\textbf{MAPS}}               & \multicolumn{3}{c}{\textbf{GuitarSet}}          & \multicolumn{3}{c}{\textbf{GAPS}}               \\ \cmidrule(l){4-15} 
\multicolumn{1}{c}{\textbf{Model}} & \textbf{\#Param}     & \multicolumn{1}{c}{\textbf{xRT}} & \textbf{OnP}  & \textbf{OnPOff} & \textbf{Frm}  & \textbf{OnP}  & \textbf{OnPOff} & \textbf{Frm}  & \textbf{OnP}  & \textbf{OnPOff} & \textbf{Frm}  & \textbf{OnP}  & \textbf{OnPOff} & \textbf{Frm}  \\ \midrule
MT3                                & 44.7M                & 68.7x                            & .958          & .833            & .890          & .801          & .460            & .732          & .891          & .780            & .883          & .659          & .304            & .616          \\
YourMT3+                           & 48.5M                & 45.6x                            & .957          & .716            & .769          & .828          & .440            & .691          & .889          & .712            & .835          & .917          & .619            & .748          \\
MuScriptor                         & 306.5M               & 17.0x                            & .766          & .337            & .625          & .806          & .426            & .683          & .808          & .469            & .732          & .781          & .530            & .742          \\
hFT-Transformer                    & 5.5M                 & 91.7x                            & .573          & .242            & .283          & .642          & .218            & .453          & .791          & .637            & .803          & .775          & .450            & .648          \\
PerceiverTF                        & 3.5M                 & 152.7x                           & .819          & .590            & .816          & .728          & .374            & .678          & .853          & .691            & .865          & .843          & .486            & .708          \\
Transkun                           & 14.1M                & 59.2x                            & .926          & .802            & .874          & .833          & .569            & .772          & .873          & .759            & .878          & .905          & .633            & .764          \\
SFT-CRNN                           & 13.2M                & 301.5x                           & .964          & .857            & .836          & .877          & \textbf{.622}   & .583          & .898          & .777            & .767          & .938          & .640            & .415          \\
Basic Pitch                        & 16.8K                & 392.7x                           & .578          & .180            & .577          & .597          & .245            & .621          & .776          & .552            & .827          & .712          & .305            & .671          \\
HPPNet-sp                          & 1.2M                 & 154.5x                           & .956          & .803            & .892          & \textbf{.882} & .611            & \textbf{.825} & .893          & .721            & .875          & .929          & .575            & .793          \\
TriAD                              & 635K                 & 256.4x                           & .934          & .772            & .879          & .863          & .563            & .794          & .889          & .736            & .874          & .921          & .581            & .782          \\
HFSFormer                          & 1.3M                 & 154.1x                           & .871          & .664            & .852          & .821          & .483            & .754          & .857          & .667            & .856          & .852          & .442            & .740          \\ \midrule
Ours, nano                         & 26.3K                & 1622.5x                          & .880          & .647            & .831          & .817          & .511            & .760          & .863          & .664            & .852          & .887          & .451            & .733          \\
Ours, small                        & 137K                 & 1335.5x                          & .928          & .770            & .881          & .848          & .525            & .778          & .881          & .724            & .869          & .915          & .577            & .777          \\
Ours, medium                       & 679K                 & 848.5x                           & .951          & .828            & .902          & .866          & .544            & .790          & .896          & .758            & .881          & .930          & .653            & .799          \\
Ours, large                        & 3.2M                 & 346.5x                           & .962          & .861            & .915          & .870          & .545            & .790          & .907          & .794            & .895          & .937          & .678            & .814          \\
Ours, x-large                      & 15.1M                & 123.0x                           & \textbf{.968} & \textbf{.878}   & \textbf{.922} & .874          & .537            & .785          & \textbf{.909} & \textbf{.803}   & \textbf{.898} & \textbf{.939} & \textbf{.685}   & \textbf{.820} \\ \bottomrule
\end{tabular}
}
\end{table*}

\begin{table*}[tb!]
\centering
\caption{Full benchmark results 2/2.}
\label{tab:full2}
\resizebox{\textwidth}{!}{%
\begin{tabular}{@{}lrrcccccccccccc@{}}
\toprule
                                   & \multicolumn{1}{l}{} & \multicolumn{1}{l}{}             & \multicolumn{3}{c}{\textbf{EGDB}}               & \multicolumn{3}{c}{\textbf{GOAT}}               & \multicolumn{3}{c}{\textbf{URMP-stem}}          & \multicolumn{3}{c}{\textbf{Slakh-stem}}         \\ \cmidrule(l){4-15} 
\multicolumn{1}{c}{\textbf{Model}} & \textbf{\#Param}     & \multicolumn{1}{c}{\textbf{xRT}} & \textbf{OnP}  & \textbf{OnPOff} & \textbf{Frm}  & \textbf{OnP}  & \textbf{OnPOff} & \textbf{Frm}  & \textbf{OnP}  & \textbf{OnPOff} & \textbf{Frm}  & \textbf{OnP}  & \textbf{OnPOff} & \textbf{Frm}  \\ \midrule
MT3                                & 44.7M                & 68.7x                            & .448          & .235            & .448          & .291          & .210            & .398          & .829          & .727            & .879          & .900          & .735            & .826          \\
YourMT3+                           & 48.5M                & 45.6x                            & .827          & .519            & .725          & .680          & .522            & .703          & .926          & .839            & .920          & .896          & .635            & .753          \\
MuScriptor                         & 306.5M               & 17.0x                            & .789          & \textbf{.635}   & .731          & .710          & .618            & .783          & .711          & .562            & .835          & .457          & .209            & .488          \\
hFT-Transformer                    & 5.5M                 & 91.7x                            & .707          & .443            & .653          & .664          & .516            & .681          & .802          & .716            & .854          & .624          & .472            & .637          \\
PerceiverTF                        & 3.5M                 & 152.7x                           & .799          & .512            & .737          & .618          & .442            & .697          & .808          & .684            & .860          & .751          & .594            & .760          \\
Transkun                           & 14.1M                & 59.2x                            & .820          & .564            & .726          & .638          & .519            & .745          & .921          & .847            & .933          & .833          & .714            & .810          \\
SFT-CRNN                           & 13.2M                & 301.5x                           & .874          & .576            & .630          & .746          & .637            & .667          & \textbf{.933} & .859            & .898          & .893          & .745            & .754          \\
Basic Pitch                        & 16.8K                & 392.7x                           & .718          & .338            & .658          & .553          & .372            & .647          & .687          & .458            & .894          & .394          & .163            & .502          \\
HPPNet-sp                          & 1.2M                 & 154.5x                           & .857          & .519            & .749          & .739          & .581            & .770          & .911          & .800            & .909          & .870          & .684            & .814          \\
TriAD                              & 635K                 & 256.4x                           & .857          & .531            & .756          & .706          & .569            & .772          & .894          & .769            & .892          & .836          & .677            & .802          \\
HFSFormer                          & 1.3M                 & 154.1x                           & .806          & .498            & .746          & .664          & .489            & .695          & .774          & .617            & .815          & .732          & .560            & .749          \\ \midrule
Ours, nano                         & 26.3K                & 1622.5x                          & .805          & .486            & .725          & .686          & .515            & .738          & .796          & .569            & .833          & .731          & .532            & .712          \\
Ours, small                        & 137K                 & 1335.5x                          & .848          & .530            & .756          & .709          & .564            & .768          & .881          & .752            & .895          & .822          & .659            & .797          \\
Ours, medium                       & 679K                 & 848.5x                           & .869          & .558            & .764          & .720          & .584            & .796          & .924          & .835            & .930          & .877          & .733            & .842          \\
Ours, large                        & 3.2M                 & 346.5x                           & .876          & .561            & .766          & .747          & .631            & .807          & .926          & .856            & .930          & .905          & .774            & .865          \\
Ours, x-large                      & 15.1M                & 123.0x                           & \textbf{.881} & .567            & \textbf{.773} & \textbf{.750} & \textbf{.638}   & \textbf{.809} & .925          & \textbf{.862}   & \textbf{.934} & \textbf{.918} & \textbf{.798}   & \textbf{.877} \\ \bottomrule
\end{tabular}
}
\end{table*}

\section{Full comparison table}
In the benchmark Table~\ref{tab:main}, OnPOff F1 score is not shown due to the limit of space. Here we show the full comparison table including this metric, in Tables~\ref{tab:full1} and \ref{tab:full2}.

\section{Implementation Details}

\smallskip\noindent\textbf{Threshold Sweep.} Many frame-level models rely on model-specific thresholds to decide whether an event is present in their frame-level probability outputs, and the chosen values have a considerable impact on performance. For the models that require such tuning, namely ours, TriAD, HPPNet-sp, hFT-Transformer, PerceiverTF, and HFSFormer, we sweep the thresholds to obtain the best performance for each. Both the onset and frame thresholds are searched over $[0.1, 0.85]$ at intervals of $0.05$, in two stages: the onset threshold is first selected to maximize OnP F1, after which the frame threshold is selected to maximize OnPOff F1 given the chosen onset threshold.

\smallskip\noindent\textbf{Batched Chunk Inference.} Passing a full song to the model at once exceeds available memory for long pieces. To bound memory consumption, we use chunked inference for all models in song-level evaluation, following the ``half'' strategy of \cite{hft}: chunks of length $l$ are processed with $l/2$ overlap, and only the central $l/2$ of each chunk contributes to the output, with these segments concatenated to form the song-level prediction. We set $l = 5$~s in all evaluations of frame-level models, matching the clip length used during training.

\section{Probing Experiment}

This section examines how the layers of the model progressively refine the CQT input toward the target piano roll. After training the medium model, we freeze its weights and train a linear probe on the output of each internal layer to predict the sustain matrix from that layer's feature map. The resulting probe accuracy indicates how linearly decodable note information is at each stage, and therefore how far the representation has been refined toward an explicit notion of notes. Results are shown in Figure~\ref{fig:probe}.

The probes reveal a clear division of labor between the block types. Already after the frontend, the probe reaches a substantial frame F1, showing that the frontend alone recovers much of the note information from the CQT. At this stage, however, considerable false-positive energy remains at harmonic positions, and the hypothesis has yet to be refined. The harmonic blocks address exactly this: they concentrate energy onto the fundamental and suppress the overtone-position responses, improving precision. The $3 \times 3$ ResNet blocks act differently. Where activations are smeared into an indistinct region, they denoise it and leave a cleaner background (layer No. 6, ResNet 1 in the Figure); where a bin stands out clearly from its neighbors, they sharpen it (ResNet 2--4), raising the confidence of notes that are latent but subthreshold until these cross the detection threshold and improve recall. Sharpening applied to the wrong bin is subsequently corrected by the following harmonic layer (e.g., layer No. 9, Harmonic 3), which finds no supporting harmonic evidence for it. The FGLSTM heads then refine temporal structure and complete the sharpening of confidence.

\end{document}